\documentclass[11pt,a4paper]{article}
\usepackage{amsmath,amssymb}
\usepackage{epsfig,graphicx}

\begin{document}

\title{\bf Massless and Massive Vector Goldstone Bosons in Nonlinear Quantum
Electrodynamics}
\author{J.~L.~Chkareuli\footnote{{\bf e-mail}:
j.chkareuli@gmail.com} ~ and ~ Z.~R.~Kepuladze\footnote{{\bf
e-mail}: kepuladze@iphac.ge}
\\
\small{\em E. Andronikashvili Institute of Physics and I.
Chavchavadze State
University} \\
\small{\em 0177 Tbilisi, Georgia} }
\date{}
\maketitle
\
\
\
\
\
\
\
\
\

\

\

\
\begin{abstract}
The spontaneous Lorentz invariance violation (SLIV) developing in
QED\ type theories with the nonlinear four-vector field constraint
\ $A_{\mu }^{2}=M^{2}$\ (where $M$ is a proposed scale of the
Lorentz violation) is considered in the case when the internal
$U(1)$ charge symmetry is also spontaneously broken. We show that
such a SLIV pattern\ induces the genuine vector Goldstone boson
which appears massless when the $U(1)$ symmetry is exact and
becomes massive in its broken phase. However, for both of phases
an apparent Lorentz violation is completely canceled out in all
the observable processes so that the physical Lorentz invariance
in theory is ultimately restored.
\end{abstract}

\

\

\

\

\

\

\

\

\

\

\

\

\
\begin{center}
{\em To appear in Proceedings of the XIVth International Seminar ``Quarks-2006''}{\it %
\ }

{\tiny 18-24 }{\it {\tiny May 2006, St-Petersburg, Russia}}
\end{center}

\thispagestyle{empty} \newpage
\section{Introduction}

Lorentz invariance and its spontaneous violation seems to play a special
role with respect to the internal local symmetries observed in particle
physics\cite{bjorken,cfn}. This violation could generally cause the
occurence of the corresponding massless Nambu-Goldstone modes which are
believed to be photons\ or other gauge fields. At the same time the
spontaneous Lorentz invariance violation (SLIV) has attracted considerable
attention in the last years as an interesting phenomenological possibility
appearing in the framework of various quantum field and string theories\cite
{alan,jakiw,glashow,moffat,mota}.

The effective theoretical laboratory for the SLIV consideration happens to
be some simple class of the QED type models for the starting massive vector
field $A_{\mu }$ where, in one way or another, the nonlinear dynamical
constraint of type
\begin{equation}
\text{\ }A_{\mu }^{2}=M^{2}\text{\ }  \label{constr}
\end{equation}
($M$ is a proposed scale of the SLIV) is appeared. This constraint means in
essence the vector field $A_{\mu }$ develops the vacuum expectation value
(VEV) and Lorentz symmetry $SO(1,3)$ formally breaks down to $SO(3)$ or $%
SO(1,2)$ depending on the sign of the $M^{2}$. Such models, from the SLIV
point of view was studied by Nambu\cite{nambu} independently of the
dynamical mechanism which could cause the spontaneous Lorentz violation. For
this purpose he applied the technique of nonlinear symmetry realizations
which appeared successful in handling the spontaneous breakdown of chiral
symmetry, particularly, as it appears in the nonlinear $\sigma $ model\cite
{GL}. It was shown, while only in the tree approximation and for the
time-like SLIV ($M^{2}>0$), that the non-linear constraint (\ref{constr})
implemented into standard QED Lagrangian containing the charged\ fermion $%
\psi (x)$%
\begin{equation}
{\cal L}_{QED}=-\frac{1}{4}F_{\mu \nu }F^{\mu \nu }+\overline{\psi }(i\gamma
\partial +m)\psi -eA_{\mu }\overline{\psi }\gamma ^{\mu }\psi \text{ \ \ }
\label{lagr1}
\end{equation}
as some supplementary condition\footnote{%
The shortest way to obtain this supplementary condition\ $A_{\mu }^{2}=M^{2}$
could be an inclusion the ``standard'' quartic vector field potential $P(A)=%
\frac{m_{A}^{2}}{2}A_{\mu }^{2}-\frac{\lambda _{A}}{4}(A_{\mu }^{2})^{2}$
into the QED Lagrangian (\ref{lagr1}) as can generally be motivated\cite
{alan} from the superstring theory. This unavoidably causes the spontaneous
breakdown of Lorentz symmetry in a regular way which goes in parallel with a
linear $\sigma $ model for pions\cite{GL}. As a result, one has a massive
Higgs mode (with mass $\sqrt{2}m_{A}$) together with massless goldstones
associated with photons. Furthermore, just as in the pion model one can go
from the linear model for the SLIV to the non-linear one taking a limit $%
\lambda _{A}\rightarrow \infty ,$ $m_{A}^{2}\rightarrow \infty $ (while
keeping the ratio $m_{A}^{2}/\lambda _{A}$ to be finite) provided that this
limit exists. This immediately leads to the constraint (\ref{constr}) for
vector potential $A_{\mu }$ with $M^{2}=\frac{\mu ^{2}}{\lambda }$, as it
appears from its equation of motion to be satisfied.} appears in fact as a
possible gauge choice which amounts to a temporal gauge for the superlarge
(as it is intuitively expected) SLIV scale $M$. At the same time, the $S$%
-matrix remains unaltered under such a gauge convention. This particular
gauge allows one \ to interpret QED in terms of the SLIV with the VEV of
vector field of the type $<A_{\mu }>_{0}$ $=(M,0,0,0)$. The SLIV, however,
is proved to be superficial as it affects only the gauge of vector potential
$A_{\mu }$ at least in the tree approximation\cite{nambu}.

Recently\cite{ac}, this result has been extended to the one-loop
approximation and for both\ the time-like ($M^{2}>0$) and space-like ($%
M^{2}<0$) Lorentz violation. All the contributions to the photon-photon,
photon-fermion and fermion-fermion interactions violating the physical
Lorentz invariance happen to be exactly cancelled with each other in the
manner observed by Nambu a long ago for the simplest tree-order diagrams.
This means that the constraint $A_{\mu }^{2}=M^{2}$ having been treated as
the nonlinear gauge choice at a tree (classical) level remains as a gauge
condition when quantum effects are taken into account as well. So, in
accordance with Nambu's original conjecture one can conclude that\ the
physical Lorentz invariance is left intact at least in the one-loop
approximation provided we consider the standard QED Lagrangian (\ref{lagr1})
(with its gauge invariant $F_{\mu \nu }F^{\mu \nu }$ kinetic term and
minimal photon-fermion coupling) taken in the flat Minkowskian space-time%
\footnote{%
For some alternative possibility see the paper\cite{cfmn}}.

We consider here the spontaneous Lorentz violation in the framework of QED
with the nonlinear four-vector field constraint (\ref{constr}) in the case
when the internal $U(1)$ charge symmetry is also spontaneously broken so
that the massless vector Goldstone boson (photon) having been generated
through the SLIV becomes then massive in the $U(1)$ symmetry Higgs phase.
For this purpose one needs to extend the starting Lagrangian ${\cal L}_{QED}$
(\ref{lagr1}) by the scalar field part
\begin{equation}
{\cal {L}}(\varphi )=|D_{\mu }\varphi |^{2}-m_{\varphi }^{2}\varphi ^{\ast
}\varphi -\frac{\lambda _{\varphi }}{2}(\varphi ^{\ast }\varphi )^{2}
\label{lagr2}
\end{equation}
where $D_{\mu }\varphi =(\partial _{\mu }-ieA_{\mu })\varphi $ is a standard
covariant derivative for the charged scalar field $\varphi $ from which the
above Goldstonic photon gets its mass. We show\ again that the apparent
Lorentz violation caused by the nonlinear SLIV constraint (\ref{constr}) is
completely canceled out in all the physical processes in the same manner as
in the massless QED case considered earlier \cite{nambu,ac}.

The paper is organized in the following way. We consider first the massive
non-linear QED Lagrangian (Sec.2) appeared once the dynamical constraint (%
\ref{constr}) is explicitly implemented into Lagrangians (\ref{lagr1}, \ref
{lagr2}) and internal $U(1)$ symmetry is spontaneously broken so that the
photon becomes massive. We derive the general Feynman rules for the basic
photon-photon and photon-fermion interactions, as well as the rules related
with Higgs sector of theory. The model appears in essence two-parametric
containing the electric charge $e$ and inverse SLIV scale $1/M$ as the
perturbation parameters so that the SLIV interactions are always
proportional some powers of them. Then in Sec.3 various SLIV processes such
the massive photon scattering off the charged fermion, Higgs boson decays
and photon-photon scattering are considered in detail. All these effects,
both in the tree and one-loop approximation, appear in fact vanishing so
that the physical Lorentz invariance is ultimately restored. And, finally,
we give our conclusions in Sec.4.

\section{\protect\bigskip The Lagrangian and Feynman rules}

\subsection{\bf \ The Lagrangian: U(1) symmetry phase}

We consider simultaneously both of the above-mentioned SLIV cases, time-like
or space-like, introducing some properly oriented unit Lorentz vector $%
n_{\mu }$ ($n_{\mu }^{2}\equiv n^{2}=\pm 1$) so as to have the following
general parametrization for the vector potential $A_{\mu }$ in the
Lagrangian (\ref{lagr1}) of the type
\begin{equation}
A_{\mu }^{2}=n^{2}M^{2},\text{\ \ \ }A_{\mu }=a_{\mu }+n_{\mu }(n\cdot A)
\label{par}
\end{equation}
(hereafter $M^{2}$ is defined strictly positive) where the $a_{\mu }$ is
pure Goldstonic mode
\begin{equation}
\text{\ }n\cdot a=0\text{\ }  \label{sup}
\end{equation}
while the Higgs mode (or the $A_{\mu }$ component in the vacuum direction)
is given by the scalar product $n\cdot A$. Substituting this parametrization
into the vector field constraint (\ref{constr}) one comes to the equation
for $n\cdot A$ (taking, for simplicity, the positive sign for the square
root and expanding it in powers of $\frac{a_{\nu }^{2}}{M^{2}}$)
\begin{equation}
\text{\ }n\cdot A\text{\ }=\left[ (M^{2}-n^{2}a_{\nu }^{2})\right] ^{\frac{1%
}{2}}=M-\frac{n^{2}a_{\nu }^{2}}{2M}+O(1/M^{2})  \label{constr1}
\end{equation}

We proceed further putting that new parametrization (\ref{par}) into our
basic Lagrangians (\ref{lagr1}) and (\ref{lagr2}), then expand it in powers
of $\frac{a_{\nu }^{2}}{M^{2}}$ and make the appropriate redefinition of the
fermion and scalar fields according to
\begin{equation}
\psi \rightarrow e^{ieM(n\cdot x)}\psi ,\text{ \ }\varphi \rightarrow
e^{ieM(n\cdot x)}\varphi
\end{equation}
so that the bilinear fermion and scalar terms, $eM\overline{\psi }(\gamma
\cdot n)\psi $ \ and $\varphi ^{\ast }[ieM(\stackrel{\Leftrightarrow }{%
\partial }\cdot n)+e^{2}n^{2}M^{2}]\varphi $, appearing, respectively, from
the expansion of the fermion and charged scalar current interactions in the
Lagrangians (\ref{lagr1}, \ref{lagr2}) are exactly cancelled by an analogous
terms stemming now from their kinetic terms (the abbreviation $\stackrel{%
\Leftrightarrow }{\partial }$ means, as usual, $\varphi ^{\ast }\stackrel{%
\Leftrightarrow }{\partial }\varphi =\varphi ^{\ast }(\partial \varphi
)-(\partial \varphi ^{\ast })\varphi $). So, we eventually arrive at the
nonlinear SLIV Lagrangian for the Goldstonic $a_{\mu }$ field (denoting its
strength tensor by $f_{\mu \nu }=\partial _{\mu }a_{\nu }-\partial _{\nu
}a_{\mu }$)
\begin{eqnarray}
{\cal L}(a,\psi ,\varphi ) &=&-\frac{1}{4}f_{\mu \nu }f^{\mu \nu }-\frac{1}{2%
}\lambda (n\cdot a)^{2}-\frac{n^{2}}{4M}f_{\mu \nu }\left[ \left( n^{\mu
}\partial ^{\nu }-n^{\nu }\partial ^{\mu }\right) a_{\rho }^{2}\right] +
\nonumber \\
&&+\overline{\psi }(i\gamma \partial +m)\psi -ea_{\mu }\overline{\psi }%
\gamma ^{\mu }\psi +\frac{en^{2}a_{\rho }^{2}}{2M}\overline{\psi }(\gamma
\cdot n)\psi +  \label{NL} \\
&&+|(\partial _{\mu }-iea_{\mu })\varphi |^{2}-\frac{ien^{2}a_{\rho }^{2}}{2M%
}[\varphi ^{\ast }(\stackrel{\Leftrightarrow }{\partial }\cdot n)\varphi ]%
\text{ }-P(\varphi )\text{ \ }  \nonumber
\end{eqnarray}
explicitly including its orthogonality condition $n\cdot a=0$ through the
term which can be treated as the gauge fixing term (taking the limit $%
\lambda \rightarrow \infty $). Note that there are presented only the terms
of the first order in $\frac{a_{\nu }^{2}}{M^{2}}$ in the Lagrangian and
also retained the former notations for the fermion $\psi $ and scalar field $%
\varphi $ (with its unchanged potential $P(\varphi )$ included, as is given
in the starting Lagrangian (\ref{lagr2})).

The Lagrangian (\ref{NL}) completes the nonlinear $\sigma $ model type
construction for quantum electrodynamics for the charged fermion and scalar
fields . The model contains the massless vector Goldstone boson modes
(keeping the massive Higgs mode frozen), and in the limit $M\rightarrow
\infty $ is indistinguishable from conventional QED taken in the general
axial (temporal or pure axial) gauge. So, for this part of the Lagrangian $%
{\cal L}(a,\psi ,\varphi )$ given by the zero-order terms in $1/M$ the
spontaneous Lorentz violation only means the noncovariant gauge choice in
otherwise the gauge invariant (and Lorentz invariant) theory. Remarkably,
furthermore, also all the other terms in the ${\cal L}(a,\psi ,\varphi )$ \ (%
\ref{NL}), though being by themselves the Lorentz and $C(CPT)$ violating
ones, cause no the physical SLIV effects which appear strictly cancelled in
all the physical processes involved. As shows the explicit calculations,
there is a full equivalence of such a model with a conventional quantum
electrodynamics at least in the tree \cite{nambu} and one-loop\cite{ac}
approximation taken for the pure fermionic part in the Lagrangian (\ref{NL}%
). The same conclusion can obviously be expected for its scalar part as
well. This seems to confirm that the starting vector field condition $A_{\mu
}^{2}=n^{2}M^{2}$ which results in the nonlinear QED model (\ref{NL}) is the
gauge choice rather non-trivial dynamical constraint which might come to the
physical Lorentz violation.

\subsection{\bf The Lagrangian: U(1) symmetry broken phase}

Let us now turn to the case of the spontaneous Lorentz violation when the
accompanying internal $U(1)$ symmetry in the SLIV Lagrangian ${\cal L}%
(a,\varphi )$ (\ref{NL}) is also spontaneously broken. For this purpose one
replaces, as usual, the scalar mass squared $m_{\varphi }^{2}\rightarrow
-m_{\varphi }^{2}$ in its potential $P(\varphi )$ so that the scalar $%
\varphi $ now develops the VEV
\begin{equation}
\varphi =\frac{1}{\sqrt{2}}(\eta (x)+v)e^{i\xi (x)/v},\text{ }%
v^{2}=2m_{\varphi }^{2}/\lambda _{\varphi }  \label{phi}
\end{equation}
where for the scalar field $\varphi $ the standard polar parametrization is
used with the proper Higgs and Goldstone modes, $\eta (x)$ and $\xi (x)$,
involved. Putting the shifted scalar field (\ref{phi}) into the Lagrangian (%
\ref{NL}) one comes to the final SLIV theory\ with the broken $U(1)$
symmetry
\begin{eqnarray}
{\cal {L}}(a,\psi ,\eta ,\xi ) &=&{\cal {L}}(a,\psi )+\frac{1}{2}(\partial
_{\mu }\eta )^{2}+\frac{1}{2}(\eta +v)^{2}[\partial _{\mu }(\xi /v)-ea_{\mu
}]^{2}+  \nonumber \\
&&+\frac{en^{2}a_{\rho }^{2}}{2M}(\eta +v)^{2}(\partial \cdot n)(\xi
/v)+P(\eta )\text{ }  \label{fin}
\end{eqnarray}
where ${\cal {L}}(a,\psi )$ stands for the vector field and fermion part
(both linear and nonlinear) as is given in the Lagrangian ${\cal L}(a,\psi
,\varphi )$ (\ref{NL}), while $P(\eta )$ denotes \ an ordinary polynomial of
the scalar Higgs component $\eta $ appeared. One can see that the vector
Goldstone boson $a_{\mu }$ acquires the mass term $\frac{1}{2}%
(e^{2}v^{2})a_{\mu }^{2}{}$. However, apart from that, there appears the
scalar-vector (goldston-goldston) mixing term in the Lagrangian ${\cal {L}}%
(a,\psi ,\eta ,\xi )$. In an ordinary Higgs mechanism case such a mixing
term can easily be removed by choosing a proper unitary gauge. However, it
is hardly possible in the SLIV case where, as is seen from the above
Lagrangian ${\cal {L}}(a,\psi ,\varphi )$ (\ref{NL}), one has already come
to the axial gauge choice for the vector Goldstonic boson $a_{\mu }$ once
the spontaneous Lorentz violation occurred. So, one may not put now extra
(unitary) gauge to get rid of the scalar Goldstone field $\xi (x)$.
Nonetheless, this field, if it were appeared in the theory, would correspond
to the unphysical particle in the sense that it could not appear as incoming
or outgoing lines in Feynman graphs, as was recently argued\cite{dams} in
the context of Standard Model taken in the axial gauge. This can be seen at
once by diagonalizing the bilinear $a-\xi $ mixing term in our Lagrangian $%
{\cal {L}}(a,\psi ,\eta ,\xi )$ (\ref{fin}) so that the $\xi $ field
disappears in it, while leading to the more complicated form for the $a$
boson gauge fixing term. In this connection, the option of an existence of
the starting $\xi $ field in the Lagrangian (\ref{fin}) , while having in
momentum space the diagonalized $a$ $\ $and $\xi $ propagators, happens to
be more convenient and transparent and we take this way in what follows.

\subsection{\bf \ The Feynman rules}

Actually, rewriting the $a-$ $\xi $ mixing term in the Lagrangian ${\cal {L}}%
(a,\psi ,\eta ,\xi )$ (\ref{fin}) in momentum space and diagonalizing it by
the substitution \
\begin{equation}
\xi (k)\rightarrow \xi (k)+i\mu \frac{k^{\mu }a_{\mu }(k)}{k^{2}}
\label{diag}
\end{equation}
where $\mu =ev$ is the vector $a$ boson mass, one has for this term
\begin{equation}
\frac{1}{2}(e\eta /\mu +1)^{2}\left[ -ik^{\mu }\xi (k)+\mu \left( \frac{%
k^{\mu }k^{\nu }}{k^{2}}-g^{\mu \nu }\right) a_{\nu }(k)\right] ^{2}
\label{bil}
\end{equation}
with the new pure $\xi (k)$ and $a_{\mu }(k)$ states appeared. Note that
just this transversal bilinear form for the $a$ boson in (\ref{bil})
together with its kinetic terms and the gauge fixing condition in the
Lagrangian ${\cal L}(a,\psi ,\varphi )$ (\ref{NL}) determines eventually the
diagonalized propagator for the massive $a$ boson of the type (in the limit $%
\lambda \rightarrow \infty $)
\begin{equation}
D_{\mu \nu }^{(a)}(k)=\frac{-i}{k^{2}-\mu ^{2}+i\epsilon }\left( g_{\mu \nu
}-\frac{n_{\mu }k_{\nu }+k_{\mu }n_{\nu }}{n\cdot k}+\frac{n^{2}k_{\mu
}k_{\nu }}{(n\cdot k)^{2}}\right)  \label{prop1}
\end{equation}
whose numerator is happened to be the same as for the axially gauged
massless vector boson. Meanwhile the propagator for the massless scalar
field $\xi $ amounts to
\begin{equation}
D^{(\xi )}(k)=\frac{i}{k^{2}}  \label{prop2}
\end{equation}
For the vector boson $a_{\mu }$ being orthogonal to $n^{\mu }$, one can
choose a basis of two transverse (in momentum space) components $\epsilon
_{\mu }^{(t)}(k)$ ($t=1,2$)
\begin{equation}
n^{\mu }\epsilon _{\mu }^{(t)}(k)=0,\text{ \ }k^{\mu }\epsilon _{\mu
}^{(t)}(k)=0  \label{trans}
\end{equation}
and the `preferred' component $\epsilon _{\mu }^{(n)}(k)$ determined by the
particular SLIV direction $n^{\mu }$
\begin{equation}
\epsilon _{\mu }^{(n)}(k)=N(k_{\mu }-n_{\mu }\frac{n\cdot k}{n^{2}}),\text{
\ }n^{\mu }\epsilon _{\mu }^{(n)}(k)=0,\text{ \ }k^{\mu }\epsilon _{\mu
}^{(n)}(k)=N(\mu ^{2}-\frac{(n\cdot k)^{2}}{n^{2}})
\end{equation}
where the normalization factor $N$ is proposed to be chosen in such way that
the sum of all the polarizations amounts to the numerator of the $a$ boson
propagator (\ref{prop1}).

Supplementing this propagator and ordinary Feynman rules by the rules
concerning the Lorentz violating interactions (see also\cite{ac}) in the
Lagrangians ${\cal {L}}(a,\psi ,\eta ,\xi )$ (\ref{fin}) and ${\cal {L}}%
(a,\psi ,\varphi )$ (\ref{NL}), particularly, those for the contact $a^{2}$%
-fermion vertex $\ \ $%
\begin{equation}
i\frac{eg_{\mu \nu }n^{2}}{M}(\gamma \cdot n)  \label{ver1}
\end{equation}
and the $a^{3}$ vertex (rewriting it first as the $-$ $\frac{n^{2}}{M}%
(\partial _{\mu }a_{\nu }n^{\mu }a_{\rho }\partial ^{\nu }a^{\rho })$)
\begin{equation}
-\frac{in^{2}}{M}\left[ (k_{1}\cdot n)k_{1,\alpha }g_{\beta \gamma
}+(k_{2}\cdot n)k_{2,\beta }g_{\alpha \gamma }+(k_{3}\cdot n)k_{3,\gamma
}g_{\alpha \beta }\right]  \label{ver2}
\end{equation}
(where the second index in the each photon 4-momentum $k_{1},$ $k_{2}$ and $%
k_{3}$ denotes its Lorentz component) one is ready to calculate some of the
low-order (in $1/M)$ processes related with the $a$ boson and fermion. Note
that the scalar field $\xi $ is not coupled to fermions and, therefore, is
not considered in the the $a$-boson-fermion interactions. However, one
should include into consideration another $a^{3}$ vertex which appears from
the $a^{2}-\xi $ coupling in the final Lagrangian ${\cal {L}}(a,\psi ,\eta
,\xi )$ (\ref{fin}) once the $a-\xi $ diagonalization (\ref{diag}) in
momentum space has been carried out:
\begin{equation}
\frac{in^{2}\mu ^{2}}{M}\left[ \frac{(k_{1}\cdot n)}{k_{1}^{2}}k_{1,\alpha
}g_{\beta \gamma }+\frac{(k_{2}\cdot n)}{k_{2}^{2}}k_{2,\beta }g_{\alpha
\gamma }+\frac{(k_{3}\cdot n)}{k_{3}^{2}}k_{3,\gamma }g_{\alpha \beta }%
\right]  \label{ver3}
\end{equation}
One can see that for the $a$ bosons being on the mass shell, $%
k_{1,2,3}^{2}=\mu ^{2}$, the vertices (\ref{ver2}) and (\ref{ver3}) exactly
cancel each other.

The other rules related with interactions of scalar Higgs and Goldstone
fields, $\eta $ and $\xi $, will be given in the next section.

\section{SLIV processes in massive QED}

We show now by a direct calculation of the tree level amplitude for Compton
scattering of the massive vector Goldstone $a$ boson off the charged fermion
and other processes that the spontaneous Lorentz violation, being
superficial in the massless nonlinear QED with an exact $U(1)$ symmetry
involved\cite{nambu,ac}, is still left hidden even though this symmetry is
spontaneously broken and the photon is getting mass.

\subsection{Vector boson scattering on fermion}

The Lorentz violating part of the elastic $a$-boson-fermion scattering is,
as follows from the Lagrangians ${\cal {L}}(a,\psi ,\eta ,\xi )$ (\ref{fin})
and ${\cal {L}}(a,\psi ,\varphi )$ (\ref{NL})), the only SLIV fermionic
process which appears in the lowest $1/M$ order. This process is concerned
with two diagrams one of which is given by the direct contact $a^{2}$%
-fermion vertex (\ref{ver1}), while another corresponds to the $a$ boson
exchange induced by the $a^{3}$ couplings (\ref{ver2}) and (\ref{ver3}).
Owing to the above-mentioned mutual cancellation of these $a^{3}$ couplings
for the on-shell $a$ bosons, only the third terms in them contributes in the
case considered so that one comes to the simple matrix element $i{\cal M}$
for the these two diagrams
\begin{equation}
i{\cal M}=i\frac{en^{2}}{M}\bar{u}(p_{2})\left[ (\gamma \cdot n)+i(1-\frac{%
\mu ^{2}}{k^{2}})(kn)k_{\beta }\gamma _{\alpha }D_{\alpha \beta }^{(a)}(k)%
\right] u(p_{1})\cdot \lbrack \epsilon (k_{1})\cdot \epsilon (k_{2})]
\label{matr1}
\end{equation}
where the spinors $u(p_{1,2})$ and polarization vectors $\epsilon (k_{1,2})$
stand for the ingoing and outgoing fermions and $a$ bosons, respectively,
while $k$ is the 4-momentum transfer $k=p_{2}-p_{1}=k_{1}-k_{2}$. After
further simplifications in the square bracket related with the explicit form
of the $a$ boson propagator $D_{\alpha \beta }^{(a)}(k)$ (\ref{prop1}) and
the fermion current conservation $\bar{u}(p_{2})(\hat{p}_{2}-\hat{p}_{1}){%
u(p_{1})=0}$, one is finally led to the total cancellation of the Lorentz
violating contributions to the Compton scattering of the massive vector
Goldstone boson $a$
\begin{equation}
i{\cal M}_{SLIV}(a+\psi \rightarrow a+\psi )=0  \label{apsi}
\end{equation}
One could say that such a result may be in some sense expected
since from the SLIV point of view the massive QED which we
considered here is hardly differed from the massless
one\cite{nambu,ac}. Actually, the fermion current conservation,
which happens crucial for the above cancellation, works in both of
cases depending no whether the internal $U(1)$ symmetry is exact
or spontaneously broken. The fermion sector (being no coupled to
the charged scalar from the outset (\ref{NL})) still possesses
this symmetry at least in tree level approximation thus leading to
the SLIV cancellation.

\subsection{Higgs boson decays}

Remarkably, the situation is not changed in the Higgs sector where the $U(1)$
symmetry related with the starting charged scalar field seems to be directly
broken and, therefore, the physical SLIV might appear. Let us examine, for
sure, the Lorentz violating Higgs boson decay $\eta \rightarrow 3a$ which
also appears in the lowest $1/M$ order if the masses of the $\eta $ and $a$
bosons are properly arranged, $m_{\eta }>3\mu $ (or $e<\sqrt{2\lambda
_{\varphi }/9}$ according to Eq.(\ref{phi})).

As one can see from the Lagrangian ${\cal {L}}(a,\psi ,\eta ,\xi )$ (\ref
{fin}) with the substitution (\ref{diag}) already made, this decay goes
through the contact $\eta -a^{3}$ coupling leading to the matrix element
\begin{equation}
i{\cal M}_{cont}=i\frac{en^{2}}{M\mu }\eta (k)\sum_{l,m,n}P^{lmn}[\epsilon
(k_{m})\cdot \epsilon (k_{n})][k_{l}\cdot \epsilon (k_{l})](k_{l}\cdot n)
\label{con}
\end{equation}
where the external 4-momenta $k_{l,m,n}$ ($l,m,n=1,2,3$) of all three $a$%
-bosons with the polarization vectors $\epsilon (k_{l,m,n})$ are supposed to
be picked up according the symmetrical projection operator $P^{lmn}$ ($%
l,m,n=1,2,3$) introduced which takes the nonzero value $1$ for only the
non-equal index values ($l\neq m\neq n$), and also the on-shell condition $%
(k_{l,m,n})_{\alpha }^{2}$ $=\mu ^{2}$ has been used; furthermore, $\eta (k)$
stands for the Higgs boson wave function and the total energy-momentum
conservation is supposed, $k=k_{l}+k_{m}+k_{n}$.

Apart from this contact diagram, the $\eta \rightarrow 3a$ decay stems via
the pole diagrams corresponding to the intermediate $a$ and $\xi $ boson
exchange. They are diagrams where the $\eta $ decays first into two $a$
bosons or into $a$ and $\xi $ bosons (with momenta $k_{1}$ and $k_{2}$) due
to the normal Lorentz invariant vertexes stemming from Eq.(\ref{bil})
\begin{eqnarray}
&&2ie\mu \left( \frac{k_{1}^{\mu }k_{1}^{\nu }}{k_{1}^{2}}-g^{\mu \nu
}\right) \left( \frac{k_{2}^{\mu }k_{2}^{\rho }}{k_{2}^{2}}-g^{\mu \rho
}\right)  \label{ita1} \\
&&2ek_{1}^{\mu }\left( \frac{k_{2}^{\mu }k_{2}^{\rho }}{k_{2}^{2}}-g^{\mu
\rho }\right)  \label{ita2}
\end{eqnarray}
followed then by the virtual Lorentz violating transitions $a\rightarrow 2a$
and $\xi \rightarrow 2a$ given, respectively, by the $a^{3}$ couplings (\ref
{ver2},\ref{ver3}) and by the $a^{2}-\xi $ vertex in the Lagrangian (\ref
{fin})
\begin{equation}
\frac{n^{2}\mu }{M}(n\cdot k)g^{\mu \nu }  \label{aksi}
\end{equation}
These six pole diagrams (three diagrams for the each type exchange)
correspond, respectively, to the cases when one of $a$ bosons with
4-momentum $k_{l}$ ($l=1;2;3$) is produced directly, whereas two other $a$
bosons with momenta $k_{m}$ and $k_{n}$ ($m,n=2,3;1,3;1,2$) appear from the
virtual $a$ and $\xi $ boson.

Using the above projection operator $P^{lmn}$ one can calculate the decay
amplitude according to all these pole diagrams simultaneously. Note that
that all the momenta in the above Feynman rules are measured ingoing so that
for the outgoing $\xi $ state the vertexes (\ref{ita2}) and (\ref{aksi})
should get a minus sign. Again, owing to the already mentioned mutual
cancellation of the $a^{3}$ vertices (\ref{ver2}) and (\ref{ver3}) for the
on-shell $a$ bosons, only one of their terms contributes in the $a$ boson
exchange diagrams. Remarkably, the non-pole contribution in these $a$ boson
exchange terms appears to be completely cancelled (when gauge fixing
condition $n\cdot \epsilon (k_{l,m,n})=0$ is used) with\ the contact diagram
contribution $i{\cal M}_{cont}$ (\ref{prop1}) , while the the pole
contribution terms are happened to be exactly cancelled with the terms
stemming from the intermediate $\xi $ boson diagrams. So, one eventually has
that the total amplitude for the Lorentz-violating $\eta \rightarrow 3a$
decay is certainly vanished
\begin{equation}
i{\cal M}_{SLIV}(\eta \rightarrow 3a)=0  \label{t}
\end{equation}

\subsection{Other processes}

In the next $1/M^{2}$ order the Lorentz violating $a-a$ scattering is also
appeared. Its amplitude is concerned with the $a$ boson exchange diagram and
the contact $a^{4}$ interaction diagram following from the higher terms in $%
\frac{a_{\nu }^{2}}{M^{2}}$ in the Lagrangian (\ref{NL}). Again, these two
diagrams are exactly cancelled giving no the physical Lorentz violating
contributions.

The same conclusion seems to be derived for the higher order processes
including both the tree diagrams and the loops concerning the $a$ bosons and
fermions. Actually, as in the massless QED case considered earlier \cite{ac}%
, the corresponding one-loop matrix elements, when they do not vanish by
themselves, amount to the differences between pairs of the similar integrals
whose integration variables are shifted relative to each other by some
constants (being in general arbitrary functions of the external four-momenta
of the particles involved) that in the framework of the dimensional
regularization leads to their total cancellation.

\bigskip

\section{\protect\bigskip Conclusions}

We have shown\ that the Lorentz violation pattern\ developing due to the
nonlinear four-vector field constraint \ $A_{\mu }^{2}=M^{2}$ in the QED
type theories induces the genuine vector Goldstone boson which appears
massless in the Coulomb phase of theory when the internal $U(1)$ charge
symmetry is exact and becomes massive in its Higgs phase once this $U(1)$
symmetry spontaneously breaks. However, for both of phases an apparent
Lorentz violation is completely canceled out in all the observable processes
so that the physical Lorentz invariance in theory is ultimately restored.
Remarkably, although the scalar Goldstone mode $\xi $ related with the
scalar field $\varphi $ is not excluded in the massive nonlinear
electrodynamics case (since one can not put the proper unitary gauge in
addition to the existing axial one (\ref{sup}) determined by the SLIV) it
does not appear as the physical particle - the pole at $k^{2}=0$ that occurs
in its propagator (\ref{prop2}) is always canceled by the poles in the
interaction vertices\ (\ref{ver2}, (\ref{ver3}) of the vector Godstone boson
$a$ emerged.

So, for the QED like theories the spontaneous Lorentz symmetry breaking
owing to the nonlinear four-vector field constraint $A_{\mu }^{2}=M^{2}$ (or
to its more familiar linearized form $A_{\mu }=a_{\mu }+n_{\mu }M$) is in
fact superficial both in massless or massive photon case even if the quantum
corrections in terms of the one-loop contributions are included into
consideration\footnote{%
Remarkably, such theories are proved to belong to some general
class of models for which the special theorem on the SLIV
non-observability appears to work\cite{cfn2}.}. This happens to
correspond only to fixing the non-covariant gauge for the vector
field in a special manner admitting an existence unphysical scalar
Goldstone mode $\xi $ in the theory provided that one starts with
an ordinary QED type model (\ref{lagr1}) with its gauge invariant
$F_{\mu \nu }F^{\mu \nu }$ kinetic term and minimal photon-matter
couplings taken in the flat Minkowskian space-time.

\section*{Acknowledgments}

We would like to thank Colin Froggatt, Rabi Mohaptra and Holger Nielsen for
useful discussions and comments. One of us (J.L.C.) is grateful for the warm
hospitality shown to him during his visit to Center for Particle and String
Theory at University of Maryland where part of this work was carried out.


\begin{thebibliography}{99}
\bibitem{bjorken}  W.~Heisenberg, Rev. Mod. Phys. {\bf 29, }269{\bf \ (}1957%
{\bf )}; J.D.~Bjorken, Ann. Phys. (N.Y.) {\bf 24, }174 (1963); T.~Eguchi,
Phys.Rev. D {\bf 14, }2755{\bf \ }(1976).

\bibitem{cfn}  J.L.~Chkareuli, C.D.~Froggatt and H.B.~Nielsen,
Phys.~Rev.~Lett.~{\bf 87, }091601 (2001); Nucl.~Phys.~B {\bf 609, }46
(2001); Per Kraus and E.T. Tomboulis, Phys. Rev. D {\bf 66, }045015 (2002);
A. Jenkins, Phys. Rev. D {\bf 69, }105007 (2004) .

\bibitem{alan}  V.A. Kostelecky and S. Samuel, Phys. Rev. D {\bf 39, }683
(1989); V.A. Kosteletsky and R. Potting, Nucl. Phys. B{\bf \ 359, }545
(1991); D.~Colladay and V.~A.~Kostelecky, Phys. Rev. D{\bf 58}, 116002
(1998); V.A. Kostelecky, Phys. Rev. D{\bf 69, }105009 (2004); R. Bluhm and
V.A. Kostelecky, Phys. Rev. D {\bf 71, }065008 (2005). For an excellent
overview of various theoretical ideas and phenomenology, see {\it CPT and
Lorentz Symmetry}, ed. A. Kostelecky (World Scientific, Singapore, 1999,
2002, 2005).

\bibitem{jakiw}  S.M. Carroll, G.B. Field and R. Jackiw, Phys. Rev. D {\bf 41%
}, 1231,1990; R. Jackiw and V.A. Kostelecky, Phys. Rev. Lett. {\bf 82},
3572, (1999).

\bibitem{glashow}  S. Coleman and S.L. Glashow, Phys. Rev. D {\bf 59, }116008%
{\bf \ }(1999).

\bibitem{moffat}  J. W. Moffat, Found. Phys. {\bf 23}, 411 (1993); Int. J.
Mod.Phys. D{\bf 2, }351 (1993); Int. J. \ \ Mod. Phys. D{\bf 12, }1279
(2003).

\bibitem{mota}  O. Bertolami and D.F. Mota, Phys. Lett. B{\bf \ 455}, 96
(1999).

\bibitem{nambu}  Y. Nambu, Progr. Theor. Phys. Suppl. Extra 190 (1968).

\bibitem{GL}  S. Weinberg, {\it The Quantum Theory of Fields,} v.2,
Cambridge University Press, 2000.

\bibitem{ac}  A.T. Azatov and J.L. Chkareuli, Phys. Rev. D {\bf 73}, 065026
(2006).

\bibitem{cfmn}  J.L.~Chkareuli, C.D.~Froggatt, R.N. Mohapatra and
H.B.~Nielsen, hep-th/0412225.

\bibitem{dams}  C. Dams and R. Kleiss, Eur. Phys. Journ. C {\bf 34, }419
(2004).

\bibitem{cfn2}  J.L.~Chkareuli, C.D.~Froggatt and H.B.~Nielsen,
hep-th/0610186, Phys.~Rev.~Lett. (to appear).
\end{thebibliography}
\end{document}